\documentclass{article}
\usepackage{amssymb}

\newcommand{\bfr}{\begin{flushright}}
\newcommand{\efr}{\end{flushright}}
 
\begin{document}
\title{Extremal Black Holes and Strings in Linear Dilaton Vacua
}
\author{Takuya Maki\\
Department of Physics, Tokyo Metropolitan University,\\
Minami-ohsawa, Hachioji-shi, Tokyo 192-03, Japan\\
and\\
Kiyoshi Shiraishi\\
Akita Junior College, Shimokitade-Sakura, Akita-shi, Akita 010, Japan
}
\date{Prog. Theor. Phys. {\bf 90} (1993) pp. 1259--1268
}
\maketitle
\begin{abstract}
Analytic solutions to low-energy string theory, which describe
arbitrary numbers of extreme `black holes (strings)' in linear
dilaton vacua, are found.
\end{abstract}

\section{Introduction}
There has recently been a great interest in the physics of Einstein
equations coupled to dilaton. Dilaton gravity arises from string
theory as a low-energy effective field theory and so exact solutions of
the theory have been constructed from the viewpoint of underlying
conformal field theories.\cite{1} (By the way, two-dimensional dilaton
gravity has also been studied intensively as a toy model in which
evaporating quantum black holes are obtained.\cite{2})

One of the most pronounced solutions to dilaton gravity closely
connected to string theory is known as a linear dilaton
solution.\cite{3} Such solutions of this type can be found in general
classes of string theory in arbitrary dimensions because only dilaton
field varies with time in the solution.

Other bosonic gauge fields such as a vector field and an
anti-symmetric tensor field are known to play important roles in
construction of macroscopic, spatially-localized object in string theory
\cite{4,10} since the dilatonic charge of the object is proportional to
the charge associated with the gauge field in general, non-trivial
configuration of the dilaton fields is realized around the localized
object. In particular, the existence of the gauge forces admits static
multisoliton solutions in string theory. Such solutions describe static
configurations of an arbitrary number of maximally-charged
(=extremal) black holes (or strings, membranes,\dots).
 
The present authors have presented the multi-soliton solutions for the
Einstein-Maxwell-dilaton system in arbitrary dimensions.\cite{11} In our
previous works, we have only treated the limited case of
electrically-charged solitons.

In the present paper, We show analytic solutions of low-energy string
theory, which describes (electrically or magnetically charged)
muti-`black hole' (`black string') configuration in the linear
dilaton vacuum.
 
Actually, our solutions have singularities which are not protected by
horizons. Here we call the solitonic solution `black hole' with
quotation marks above, since the special case with a single soliton
corresponds to the extremal limit of a charged black hole in the limit
of the vanishing cosmological constant.\cite{6,10,11} We will omit the
quotation marks hereafter simply for brevity.
 
The outline of the present paper is the following. We treat several
different models and solitonic solutions in parallel sections
$2\sim 5$. In
\S 2, we obtain the multi-magnetic black hole solutions in the
low-energy string theory including a cosmological constant. The
electric black hole solution with linear dilaton background is
constructed in \S 3. The difference between electric and magnetic
solutions is emphasized. Section 4 contains the description of
multi-black hole solution with a non-zero three form field strength.
The solution describing multi-black strings with linear dilaton is
treated in \S 5. The final section is devoted to a brief discussion.

\section{Magnetic black hole solution with linear dilaton in four
dimensions}
In this section, we consider multi-black hole solution in four
dimensions. The low-energy heterotic string effective action for the
bosonic fields in $D$ dimensions can be written as
\begin{equation}
S=\int d^Dx \frac{\sqrt{-g}}{16\pi}e^{-2\phi}[R+4(\nabla\phi)^2-F^2
-\Lambda]\,,  
\label{2.1}
\end{equation}
where $\Lambda$ is a `cosmological' constant depending on the
spacetime dimensions and the central charge of a conformal field theory
coupled to the string sigma model action. We must note that the
`cosmological constant' is coupled to the dilaton field $\phi$. Here we
omit the term involving antisymmetric field strength as well as vector
gauge field strengths except for $U(1)$ Maxwell field strength, $F$,
because they are irrelevant for the classical solution we will consider
in this section.
 
The equations of motion following from the action (\ref{2.1}) are
\begin{eqnarray}
&
&R_{\mu\nu}+2\nabla_\mu\nabla_\nu\phi-2F_{\mu\lambda}F_\nu{}^{\lambda}
=0\,, \label{2.2a}\\
& &\partial_\mu(\sqrt{-g}e^{-2\phi}F^{\mu\nu})=0\,,                   
\label{2.2b}\\ & &
R+4\nabla^2\phi-4(\nabla\phi)^2-F^2-\Lambda=0\,.
\label{2.2c}
\end{eqnarray}

If we choose the Minkowski spacetime, a solution to
(\ref{2.2a},\ref{2.2b},\ref{2.2c}) for
$\Lambda>0$ is
\begin{equation}
\phi=\phi_0(t)=\frac{\sqrt{\Lambda}}{2}t
\label{2.3a}
\end{equation}
and
\begin{equation}
F=0\,.
\label{2.3b}
\end{equation}

Here we select spatially-homogeneous evolution of the dilaton field.
This classical background is known as an exact solution to string
theory and called a linear dilaton solution.\cite{3}
 
Now we derive analytic solutions to (\ref{2.2a},\ref{2.2b},\ref{2.2c}) 
describing an arbitrary number of magnetic black holes in the linear
dilaton vacua in four dimensions. We adopt the magnetic field created
by the magnetic charge $Q_i$ located at ${\bf x}_i$,
which is written as:
\begin{equation}
F_{jk}=\varepsilon_{jkl}\sum_i \frac{Q_i}{|{\bf x}-{\bf x}_i|^2}
\frac{{\bf x}^l-{\bf x}_i^l}{|{\bf x}-{\bf x}_i|}\,,
\label{2.4}
\end{equation}
where $\varepsilon_{jkl}$ is a totally-antisymmetric tensor 
($j, k, l=1, 2, 3$).

For $\Lambda=0$, the multi-black hole solution takes the form, as noted
in Ref.~\cite{6}:
\begin{eqnarray}
& &ds^2=-dt^2+U^4({\bf x})d{\bf x}^2\,,
\label{2.5}\\
& &e^{2\phi}=U^2({\bf x})
\label{2.6}
\end{eqnarray}
with
\begin{equation}
U^2({\bf x})=1+\sum_i\frac{2M_i}{|{\bf x}-{\bf x}_i|}\,,          
\label{2.7}
\end{equation}
where $M_i=|Q_i|/\sqrt{2}$.

One can easily find a time-dependent solution to
(\ref{2.2a},\ref{2.2b},\ref{2.2c})  with $\Lambda\ne 0$ as an extension
of the static one in this case. A simple solution turns out to be
\begin{equation}
e^{2\phi}=e^{2\phi_0(t)}U^2({\bf x})\,,
\label{2.8}
\end{equation}
and the metric takes the same form as (\ref{2.5}) with the same
definition of $U^2$ (\ref{2.7}). $\phi_0(t)$ is given by (\ref{2.3a}).
This result exhibits a very simple mergence of linear dilaton and
multi-black hole solutions.

In the next section we will investigate an electrically charged black
hole solution. Electrically charged solutions may have non-trivial
dependence on time because the equations of motion are not invariant
under a duality rotation \cite{4,6} if $\Lambda\ne 0$.

\section{Electric black hole solution with linear dilaton in arbitrary
dimensions}
In this section we write down analytic solutions describing
electrically charged extreme black holes in a linear dilaton vacuum
in $D$-dimensional low-energy string theory. The equations of
motion for metric, $U(1)$ vector field, and dilaton field are the same
as (\ref{2.2a},\ref{2.2b},\ref{2.2c}).

For $\Lambda=0$ and $D=4$, the electrically-charged multisoliton
solution can be derived from the magnetic solution 
(\ref{2.5},\ref{2.6},\ref{2.7}) using a duality rotation.\cite{4,6} The
electric solution is given by
\begin{eqnarray}
& &ds^2=-U^4({\bf x})dt^2+d{\bf x}^2\,,
\label{3.1}\\
& &e^{-2\phi}=U^2({\bf x})\,,
\label{3.2}\\
& &A_\mu dx^\mu=U^{-2}({\bf x})\sum_i\frac{Q_i}{|{\bf x}-{\bf x}_i|}
dt
\label{3.3}
\end{eqnarray}
with
\begin{equation}
U^2({\bf x})=1+\sum_i\frac{2M_i}{|{\bf x}-{\bf x}_i|}\,,
\label{3.4}
\end{equation}
where $M_i=|Q_i|/\sqrt{2}$.

For arbitrary $D$, the solution can be expressed as \cite{10}
\begin{eqnarray}
& &ds^2=-U^4({\bf x})dt^2+d{\bf x}^2\,,
\label{3.5}\\
& &e^{-2\phi}=U^2({\bf x})\,,
\label{3.6}\\
& &A_\mu dx^\mu=U^{-2}({\bf x})\sum_i\frac{Q_i}{(D-3)|{\bf x}-{\bf
x}_i|^{D-3}} dt
\label{3.7}
\end{eqnarray}
with
\begin{equation}
U^2({\bf x})=1+\sum_i\frac{\mu_i}{(D-3)|{\bf x}-{\bf x}_i|^{D-3}}\,,
\label{3.8}
\end{equation}
where $d{\bf x}^2=\sum_{k=1}^{D-1}dx^kdx^k$, and $\mu_i=\sqrt{2}|Q_i|$.
$(D-1)$-dimensional space is completely flat in this solution.
  
Now we turn to the case with non-zero $\Lambda$ in arbitrary dimension.
Judging from the form of the solution with $\Lambda=0$, we naturally
take the following simplest ansatz for the metric and the dilaton field:
\begin{eqnarray}
& &ds^2=-U^{-4}({\bf x}, t)dt^2+d{\bf x}^2\,,
\label{3.9}\\ & &e^{-2\phi}=e^{-2\phi_0(t)}U^2({\bf x},
t)\,,\label{3.10}
\end{eqnarray}
where $\phi_0(t)$ is given by (\ref{2.3a}). This ansatz can
automatically connect the multisoliton solution with $\Lambda=0$ and
the linear dilaton solution with Minkowski spacetime.
  
For the electric field, one can get the general solution to
(\ref{2.2b}), which stands for the collection of the static point
charges. This takes the form as 
\begin{equation}
e^{-2\phi}U^{-2}F^{tk}=\sum_i\frac{Q_i}{|{\bf x}-{\bf x}_i|^{D-2}}
\frac{{\bf x}-{\bf x}_i}{|{\bf x}-{\bf x}_i|}\,,
\label{3.11}
\end{equation}
where $Q_i$ is the electric charge of each black hole.
 
We then substitute (\ref{3.9},\ref{3.10},\ref{3.11}) into
Eq.~(\ref{2.2a},\ref{2.2b},\ref{2.2c}). $[0, i]$ components of
(\ref{2.2a}) require
\begin{equation}
\frac{\partial^2}{\partial t\partial x^i}(e^{-2\phi_0(t)}U^2({\bf x},
t))=0\,.
\label{3.12}
\end{equation}

By taking the explicit expression (\ref{3.11}) into consideration, we
find 
\begin{equation}
U^2({\bf x}, t)=1+\sum_i\frac{\mu_i e^{2\phi_0(t)}}{(D-3)|
{\bf x}-{\bf x}_i|^{D-3}}\,,
\label{3.13}
\end{equation}
where $\mu_i=\sqrt{2}|Q_i|$.
 
We find this $U^2$ satisfies the rest of Eqs.~(\ref{2.2b}) and
(\ref{2.2c}). Then the vector field is expressed as
\begin{equation}
A_t=\pm\frac{1}{\sqrt{2}}(1-U^{-2}({\bf x}, t))\,.
\label{3.14}
\end{equation}

Now we get the electric black hole solution with a linear dilaton.%
\footnote{We have also examined more general ansatze
for $g_{\mu\nu}$ and $e^{-2\phi}\approx(f(t))^p(U^2({\bf x}, t))^q$
obtained no other solution in closed form.}
  
The difference from the magnetic solution is obvious. For the
electric black holes. the mass of each black hole seems to increase
(decrease) as the value of the linear dilaton field becomes large
(small) proportionally to time. The electrically charged solution
approaches the simple linear dilaton vacuum \cite{3} either in past or
in future infinity.

In the rest of this section, we investigate the multisoliton solutions
with the time-dependent dilaton field in a modified model.

In order to analyze the role of dilaton, one can consider an arbitrary
value for the dilaton coupling. To simplify the analysis, we first
rescale the metric by defining
\begin{equation}
\tilde{g}_{\mu\nu}=e^{-4\phi/(D-2)}g_{\mu\nu}\,.                     
\label{3.15}
\end{equation}

The action then becomes
\begin{equation}
\tilde{S}=\int d^D\tilde{x}\frac{\sqrt{-g}}{16\pi}\left[\tilde{R}-
\frac{4}{D-2}(\tilde{\nabla}\phi)^2-e^{-4\phi/(D-2)}\tilde{F}^2-
e^{4\phi/(D-2)}\Lambda\right]\,.
\label{3.16}
\end{equation}
Here the action contains the standard Einstein-Hilbert action. To treat
general dilaton couplings, we consider the action
\begin{equation}
\tilde{S}=\int d^D\tilde{x}\frac{\sqrt{-g}}{16\pi}\left[\tilde{R}-
\frac{4}{D-2}(\tilde{\nabla}\phi)^2-e^{-4a\phi/(D-2)}\tilde{F}^2-
e^{4a\phi/(D-2)}\Lambda\right]\,,
\label{3.17}
\end{equation}
where $a$ represents dilaton couplings to other entities. Obviously,
(\ref{3.17}) is reduced to (\ref{3.16}) if $a=1$.
  
One can find analytic solutions to the field equations derived from
(\ref{3.17}) for all $a$. For non-zero values for $a$ and $\Lambda$, the
set of solutions takes the form:
\begin{eqnarray}
& &d\tilde{s}^2=-U^{-4(D-3)/(D-3+a^2)}({\bf x}, \tilde{t})d\tilde{t}^2+
(\tilde{t}/t_0)^{2/a^2}U^{4/(D-3+a^2)}({\bf x}, \tilde{t})d{\bf x}^2\,,
\label{3.18}\\
&
&e^{-4a\phi/(D-2)}=(\tilde{t}/t_0)^2U^{4a\phi/(D-3+a^2)}\,,                
\label{3.19}\\ & &A_\mu d\tilde{x}^\mu=\sqrt{\frac{D-2}{2(D-3+a^2)}}
\frac{t_0}{\tilde{t}}(1-U^{-2}({\bf x}, \tilde{t}))d\tilde{t}\,,    
\label{3.20}
\end{eqnarray}
where
\begin{equation}
U^2({\bf x},
\tilde{t})=1+\sum_i\frac{\mu_i}{(\tilde{t}/t_0)^{D-3+a^2}(D-3)|{\bf x}-
{\bf x}_i|^{D-3}}
\label{3.21}
\end{equation}
and
\begin{equation}
t_0^2=\frac{(D-2)(D-1-a^2)}{\Lambda a^4}\,.
\label{3.22}
\end{equation}

For $\Lambda=0$, the static solution has been shown in Ref.~\cite{10}.
If we set $\tilde{t}=t_0$ in (\ref{3.18},\ref{3.19},\ref{3.20}), these
expressions coincide with those of the static solution. For $a=0$ and
$\Lambda\ne 0$, the exact solution describing extreme black holes in de
Sitter background is known.\cite{12} For $a=0$ and $\Lambda=0$, the
solution is, of course, reduced to the well-known
Papapetrou-Majumdar-Myers solution.\cite{13}
  
We also find that the magnetic multisoliton solution in this modified
rnodel with $\Lambda\ne 0$ can be obtained in closed form only for
$a=1$, as long as the components of metric and the exponent of the
dilaton field is written in the form $\approx(f(t))^p(U^2({\bf x},
t))^q$. The solution for
$a=1$ is of course reduced to the solution in \S 2 after the Weyl
transformation and a transformation of the time coordinate.

\section{`Magnetic' black hole solution with linear dilaton in five
dimensions}
In this section, we consider the multi-black hole solution in five
dimensions. The low-energy string effective action for the bosonic
fields in $D$ dimensions can be written as
\begin{equation}
S=\int d^Dx
\frac{\sqrt{-g}}{16\pi}e^{-2\phi}\left[R+4(\nabla\phi)^2-\frac{1}{12}
H^2-\Lambda
\right] ,
\label{4.1}
\end{equation}
where $H_{\mu\nu\lambda}=\partial_\mu B_{\nu\lambda}+
\partial_\nu B_{\lambda\mu}+\partial_\lambda B_{\mu\nu}$ is an
antisymmetric tensor field strength, which appears commonly in string
theory.

The equations of motion derived from (\ref{4.1}) are
\begin{eqnarray}
&
&R_{\mu\nu}+2\nabla_\mu\nabla_\nu\phi-\frac{1}{4}H_{\mu\lambda\sigma}
H_\nu{}^{\lambda\sigma}=0\,,
\label{4.2a}\\ & &\partial_\mu(\sqrt{-g}e^{-2\phi}H^{\mu\nu\lambda})
=0\,,
\label{4.2b}\\ & &R.+4\nabla^2\phi-4(\nabla\phi)^2-\frac{1}{12}H^2
-\Lambda=0\,.
\label{4.2c}
\end{eqnarray}

In this section, we consider the case $D=5$, as the simplest case. The
equation of motion (\ref{4.2b}) for $H$ can be satisfied by setting
\begin{equation}
H_{jkl}=\varepsilon_{jklm}\sum_i\frac{Q_i}{|{\bf x}-{\bf x}_i|^2}
\frac{{x}^m-{x}_i^m}{|{\bf x}-{\bf x}_i|}\,,
\label{4.3}
\end{equation}
where $j, k, l, m, \cdots$ run over $1, 2, 3, 4$. This stands for an
arbitrary number of charges $Q_i$ located at ${\bf x}_i$. The
configuration is often called `magnetic' field, for the suffices of the
field strength run only over spatial indices.

The solutions to Eq.~(\ref{4.2a},\ref{4.2b},\ref{4.2c}) with the charge
configuration (\ref{4.3}), which describe a single extremal and
non-extremal black hole, have been studied in Refs.~\cite{4}, \cite{5}
and \cite{7} for various $D$.

A solution to (\ref{4.2a},\ref{4.2c}) with (\ref{4.3}) in five
dimensions can easily be found and written in the following form:
\begin{equation}
ds^2=-dt^2+U^2({\bf x})d{\bf x}^2
\label{4.4}
\end{equation}
with
\begin{equation}
e^{2\phi}=e^{2\phi_0(t)}U^2({\bf x})
\label{4.5}
\end{equation}
with
\begin{equation}
U^2({\bf x})=1+\sum_i\frac{|Q_i|}{2\sqrt{2}|{\bf x}-{\bf
x}_i|^2}\,,
\label{4.6}
\end{equation}
where $\phi_0$ is given by (\ref{2.3a}).
  
This solution seems much alike the magnetic solution obtained in \S 2.
The metric is independent of the value of the cosmological
constant $\Lambda$. One can find similar exact solutions to the
equations derived from the action which contains  a $(D-2)$-form
`magnetic' field strength (instead of $H$) in $D$ dimensions. The
effect of the linear dilaton enforces $\phi\rightarrow\phi+\phi_0$ in
such solutions while the metric is unchanged.

\section{Black string solution with linear dilaton in arbitrary
dimensions}
Black string solutions were found and investigated in Refs.~\cite{4},
\cite{8} and \cite{9}. In particular, extremal black strings turn out to
be viewed as straight fundamental strings. 
In this section,
we exhibit multi-black string solutions in the linear dilaton vacuum in
$D$ dimensions.
 
The equations of motion are the same ones as
(\ref{4.2a},\ref{4.2b},\ref{4.2c}). We assume that all the strings lie
along the $z$-direction, which is perpendicular to other $(D-2)$
dimensions spanned by the coordinates $x^i$. The solution is also
assumed to be translationally invariant along the $z$-direction. We
adopt an ansatz on the three form field strength such that
\begin{equation}
\sqrt{-g}e^{-2\phi}H^{tzk}=\sum_i\frac{Q_i}{|{\bf x}-{\bf x}_i|^{D-3}}
\frac{{x}^k-{x}_i^k}{|{\bf x}-{\bf x}_i|}\,,
\label{5.1}
\end{equation}
where $Q_i$ is the `electric' charge of each string.
 
By an analysis similar to the ones in the previous sections, one
obtains the
solution describing extremal black strings in the linear dilaton vacuum.
It is given by the expression:
\begin{eqnarray}
& &ds^2=-U^{-2}({\bf x}, t)(-dt^2+dz^2)+d{\bf x}^2\,,
\label{5.2}\\
& &e^{-2\phi}=e^{-2\phi_0(t)}U^2({\bf x}, t)\,,
\label{5.3}\\
& &B_{zt}=\pm(1-U^{-2}({\bf x}, t))
\label{5.4}
\end{eqnarray}
with
\begin{eqnarray}
& &U^2({\bf x}, t)=1+\sum_i\frac{|Q_i|e^{2\phi_0(t)}}{(D-4)|{\bf
x}-{\bf x}_i|^{D-4}}\quad\mbox{for } D>5\,,    
\label{5.5a} \\& &U^2({\bf x}, t)=1-\sum_i|Q_i|e^{2\phi_0(t)}\ln(|{\bf
x}-{\bf x}_i|)\quad\mbox{for } D=4\,.
\label{5.5b}
\end{eqnarray}
If we send the value of $\Lambda$ to zero in this solution, we get the
static configuration of multi-black strings.
 
We found no other analytic solution for $\Lambda\ne 0$ in modified
models where the dilaton coupling takes a different value after scaling
similarly to (\ref{3.15}).
  
For $\Lambda=0$, the modified model is written as, for example,
\begin{equation}
\tilde{S}=\int d^D\tilde{x}\frac{\sqrt{-\tilde{g}}}{16\pi}
\left[\tilde{R}-\frac{4}{D-2}(\nabla\phi)^2-\frac{1}{12}
e^{-8a\phi/(D-2)}\tilde{H}^2\right]\,,
\label{5.6}
\end{equation}
where $a$ represents dilaton couplings to other entities. The action
(\ref{5.6}) can be connected to the action (\ref{4.1}) with $\Lambda=0$
by the Weyl transformation
 (\ref{3.15}), only if $a=1$.

An analytic solution to the field equations derived from (\ref{5.6})
for an arbitrary value for $a$ takes the form ($D>5$):
\begin{eqnarray}
& &d\tilde{s}^2=U^{-2(D-4)/(D-4+2a^2)}({\bf x})(-d\tilde{t}^2+dz^2)+
U^{4/(D-4+2a^2)}({\bf x}) d{\bf x}^2
\label{5.7}\\ & &e^{-4a\phi/(D-2)}=U^{4a^2/(D-4+2a^2)}\,,
\label{5.8}\\
& & B_{zt}=\pm\sqrt{\frac{D-2}{D-4+2a^2}}(1-U^{-2}({\bf x}))\,,
\label{5.9}
\end{eqnarray}
where
\begin{equation}
U^2({\bf x})=1+\sum_i\frac{\mu_i}{(D-3)|{\bf x}-{\bf x}_i|^{D-3}}\,.
\label{5.10}
\end{equation}

The static solution for $a=1$ is equivalent to the solution
(\ref{5.2},\ref{5.3},\ref{5.4},\ref{5.5}) with $\Lambda=0$ after the
Weyl transformation and a transformation of the time coordinate are
done.

\section{Discussion}
We have exhibited multisoliton solutions in string theory with a
cosmological term, which represent extremal black holes (strings) in
linear dilaton vacua. The metric for the electrically-charged solution
has time dependence while the one of the magnetically-charged object
is constant in time. The duality symmetry which exists in the theory
with vanishing cosmological constant is clearly broken.

It is expected that the interactions between solitonic objects in string theory with
non-zero cosmplogical constant will be explored by further study on
these solutions.
 
Here we briefly discuss on two topics connected to string theory below.
 
First we investigate the tachyon in the background spacetime expressed
by the solution we have obtained. The linearized equation for tachyon
field $T$ is given by
\begin{equation}
\frac{1}{\sqrt{-g}}\partial_\mu(e^{-2\phi}\sqrt{-g}g^{\mu\nu}
\partial_\nu T)+\mu^2 e^{-2\phi}T=0\,,
\label{6.1}
\end{equation}
where $-\mu^2$ denotes the tachyon mass. One can find that spatially
homogeneous solutions for $T(t)$ are obtained in the background of
magnetic black hole spacetime in four dimensions (in \S 2) and in
five dimensions (in \S 4). The solution is then written as
\begin{equation}
T(t)\approx e^{pt}\,,\quad\mbox{where }
p=\partial_t\phi_0\pm\sqrt{(\partial_t\phi_0)^2+\mu^2}\,.
\label{6.2}
\end{equation}

This result is independent of the existence of black holes and thus
is the known result: for example, in bosonic string theory, 
$\Lambda=(D-26)/3$ and $\mu^2=2$ ($\alpha'=2$), thus we find
$p=(\sqrt{D-26}\pm\sqrt{D-2})/\sqrt{12}$. For electric
black hole and black string spacetime, there is no homogeneous
solution of the tachyon field. This is due to the non-trivial $g_{tt}$
as well as the time-dependence of
$U^2$.
  
The relation between the black string solution and plane-fronted waves
is known.\cite{4,9} We briefly investigate the relation in terms of
our solution with a linear dilaton. In string theory, a discrete
symmetry is known for the background field. If the metric $g_{\mu\nu}$,
dilaton field $\phi$, and antisymetric field $B_{\mu\nu}$ are
independent of the coordinate $z$, there is a dual solution
\cite{4,9,14}
\begin{eqnarray}
& &\hat{g}_{zz}=1/g_{zz}\,,\quad \hat{g}_{z\alpha}=B_{z\alpha}/g_{zz}\,,
\label{6.2a}\\ &
&\hat{g}_{\alpha\beta}=g_{\alpha\beta}-(g_{z\alpha}g_{z\beta-
B_{z\alpha}B_{z\beta}})/B_{zz}\,,\label{6.2b}\\ &
&\hat{B}_{z\alpha}=g_{z\alpha}/g_{zz}\,,\quad
\hat{B}_{\alpha\beta}=B_{\alpha\beta}-2g_{z[\alpha}B_{\beta]z}/g_{zz}
\,,\label{6.2c}\\
& &\hat{\phi}=\phi-\frac{1}{2}\ln g_{zz}\,,
\end{eqnarray}
where $\alpha, \beta$ run over all directions except $z$.
Although our solution obtained in \S 5 is
the one to low-energy field equation, the dual transformation
(\ref{6.2}) should give at least an approximation to the exact solution
in string theory, which is worth studying. The dualization of the
solution (\ref{5.2},\ref{5.3},\ref{5.4}) for $D\ge 5$ yields
\begin{eqnarray}
&
&ds^2=-\left(1-\sum_i\frac{|Q_i|e^{2\phi_0(t)}}{(D-4)|{\bf
x}-{\bf
x}_i|^{D-4}}\right)dt^2+\sum_i\frac{2|Q_i|e^{2\phi_0(t)}}{(D-4)|{\bf
x}-{\bf x}_i|^{D-4}}dtdz\nonumber \\ &
&+\left(1+\sum_i\frac{|Q_i|e^{2\phi_0(t)}}{(D-4)|{\bf
x}-{\bf
x}_i|^{D-4}}\right)dz^2+d{\bf x}^2\,,
\label{6.3a}\\ &
&\phi=\phi_0(t)\,,\quad B=0\,.
\label{6.3b}
\end{eqnarray}

Using new coordinates $z=(u-v)/2$ and $t=(u+v)/2$, we can express
the solution as
\begin{equation}
ds^2=-dudv+d{\bf
x}^2+\sum_i\frac{|Q_i|e^{2\phi_0((u+v)/2)}}{(D-4)|{\bf x}-{\bf
x}_i|^{D-4}}du^2
\label{6.4}
\end{equation}
with a dilaton field $\phi_0((u+v)/2)$. This is not a precise
`plane'-fronted wave due to the dilaton contribution. This is an
eEtrernely impressive expression. We guess the metric has a physical
application in noncritical string theory.
  
We also believe that it is worth while to investigate whether there is
exact conformal field theories whose low-energy limit corresponding to
the multisoliton solutions obtained in the present paper.



\bigskip
\noindent
{\bf Note added:} While this paper was refreed, we received the paper
\cite{15} which treated the same solution in different coordinates.
\end{document}